# Dephasing due to *Which Path* Detector

E. Buks, R. Schuster, M. Heiblum, D. Mahalu, and V. Umansky

*Braun Center for Submicron Research, Department of Condensed Matter Physics,*

*Weizmann Institute of Science, Rehovot 76100, Israel*

**Abstract**

We study dephasing of electrons induced by a *which path* detector and thus verify Bohr's complementarity principle for fermions. We utilize a *double path* interferometer with two slits, with one slit being replaced by a coherent *quantum dot* (QD). A short one dimensional channel, in the form of a *quantum point contact* (QPC), in close proximity to the QD, serves as a *which path* detector. We find that by varying the properties of the QPC detector we affect the visibility of the interference, inducing thus dephasing. We develop a simple model to explain the dephasing due to the nearby detector and find good agreement with the experiment.

Bohr's complementarity principle excludes the possibility of observation of multiple paths interference simultaneously with *which path* (WP) determination [1]. Wave - like behavior, or interference, is possible only when the different possible paths a particle can take are indistinguishable, even in principle, therefore, their wave functions add coherently and thus interfere. Determining the actual path, by coupling to a *measuring* environment results in *dephasing*, namely, suppression of interference. Such principles have been demonstrated recently using parametric down conversion of photons where one photon had been used to determine the path of the second, thus preventing single photon interference [2]. Mesoscopic systems [3] can be used as ideal tools to study the interplay between interference and dephasing. Observation of quantum interference in such systems requires the absence of dephasing, prevalent due to interaction between the interfering electrons and the environment (other electrons, phonons, etc.) [4]. Nano fabrication and low temperature techniques allow observation of a variety of coherent effects, such as Aharonov Bohm (AB) interference, weak localization, resonant tunneling and conductance quantization [3]. In the present work we study a new dephasing mechanism induced by an artificial and controllable WP detector.

We utilize in the experiment an electronic *double path* interferometer [5, 6], fabricated in the plane of a high mobility two dimensional electron gas (2DEG). The two paths are defined by two slits electrons can pass through. One slit is in the form of a quantum dot (QD) [7] and the other is a *quantum point contact* (QPC). We use a coherent QD for one slit in order to allow the electronic partial wave, while staying coherent, to dwell long enough near the detector as it goes through this slit, so it can be detected more easily. Nearby the QD, but electrically separated from it, a QPC is fabricated, serving as a WP detector. It is expected that an electron passing through the QD - slit will interact with the nearby QPC - detector and modify its conductance [8]. Note that even though there is no tunneling between the interferometer and the QPC - detector, the two systems are entangled due to their mutual interaction [9]. The dephasing induced by this entanglement is studied via measuring the visibility of the AB conductance oscillations [10] produced by the double path interferometer (defined as the ratio between the peak - to - peak value of the AB oscillations and twice the average conductance). We determine experimentally the dependence of the visibility on the transmission probability of the QPC - detector and the rate electrons probe it, both determined by gate voltage and drain - source voltage across the detector, respectively. We also derive a simple theory that agrees with recent theories and with our experimental results.

The WP interferometer, seen in Fig. 1, consists of a patterned high mobility 2DEG (with density $n_s = 3.0 \cdot 10^{11}$ cm$^{-2}$ and low temperature mobility $\mu = 2.8 \cdot 10^6$ cm$^2$/Vs) formed 60 nm below the surface of a GaAs-AlGaAs heterostructure. The potential barriers and the openings in the plane of the 2DEG are induced by negatively biased, with respect to the 2DEG, miniature metal gates deposited on the surface of the heterostructure, thus depleting the electrons underneath the gates. Figure 1(b) shows an electron micrograph of the device's surface. The *two path* interferometer (see Fig. 1(a)) consists of emitter $E$ and collector $C$ constrictions, each formed by a single mode QPC, and base region $B$ in between. The grounded base contacts ($V_B = 0$) serve as draining reservoirs for scattered electrons, ensuring that only the two forward paths reach the detector. The emitter is separated from the collector by a barrier with two openings (slits). The left slit is in a form of a QPC with a metallic air bridge above connecting the central metal island. The right slit is a QD . Another QPC on the right side of the QD - slit, serving as a WP detector, is a part of a separated electronic circuit. Since a collector signal is usually too small to be measured in a totally open configuration (without reflecting barriers) [6], we incorporated additional barriers (the white gates in Fig. 1(a)) to direct the emitted electrons from the emitter into the two slits and subsequently to guide them towards the collector. However, openings between the gates still allow reflected electrons to be collected by the base, assuring thus that the collector signal is made only of the two forward propagating paths (rather than many paths that circle around a closed interferometer). The system is being cooled in a dilution refrigerator with an electron temperature $\Theta \approx 80$ mK. All measurements are done with an *ac* excitation voltage $V_E = 10\ \mu$V applied across the QPC that form the emitter injector. Under these conditions both the phase coherence length and the elastic mean free path of the electrons exceed the entire size of the interferometer.

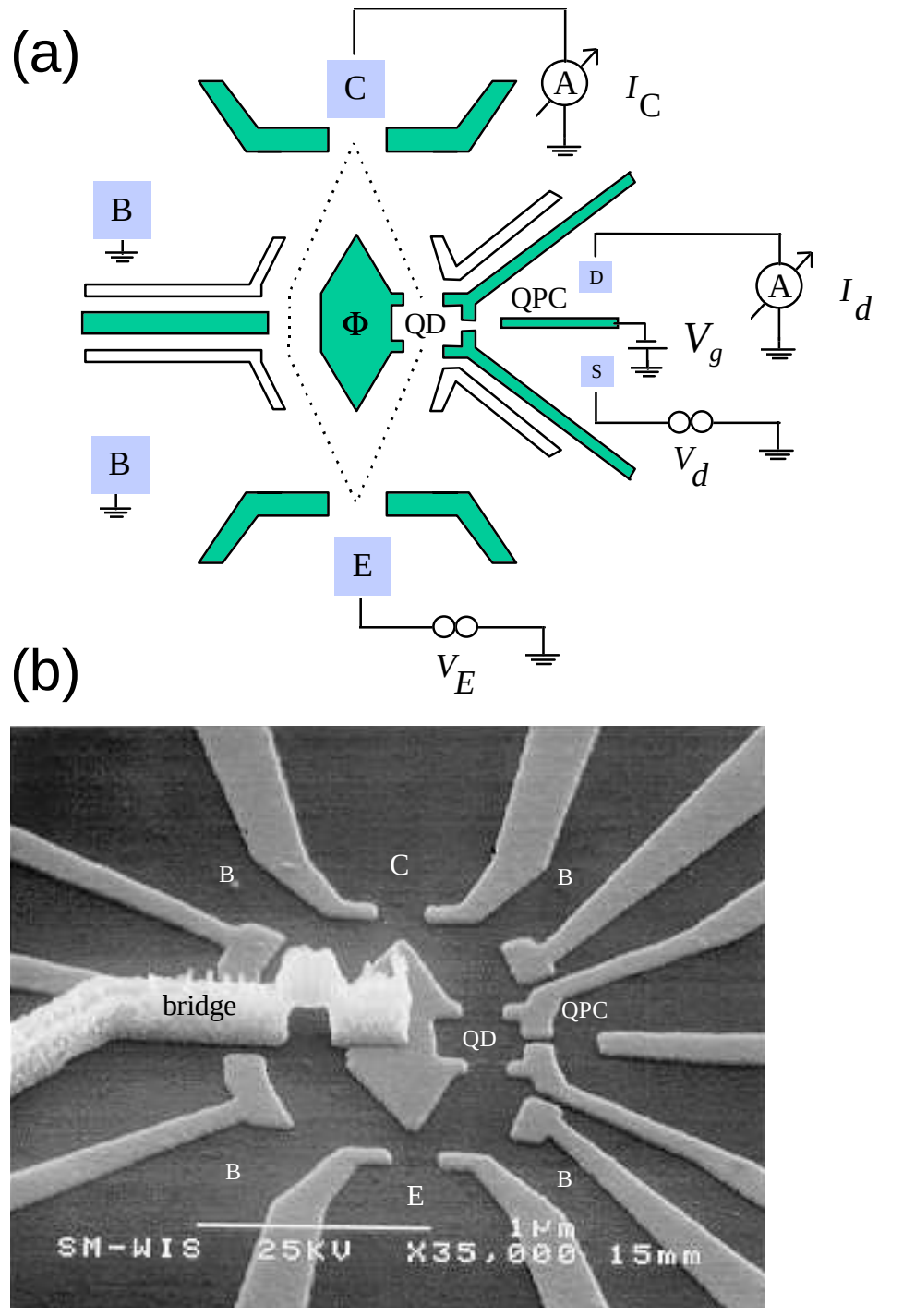

**Figure 1**: (a) A schematic description of the top electrodes and contacts of the interferometer and the detector. The interferometer is composed of three different regions, emitter $E$, collector $C$, and base regions $B$ on both sides of the barrier with the two slits. The right slit is in a form of a QD (with area $0.4 \times 0.4\ \mu\text{m}^2$) with a QPC on its right side serving as a WP detector. (b) A top view SEM micrograph of the device. The gray areas are metallic gates deposited on the surface of the heterostructure. A special lithographic technique, involving a metallic air bridge, is used to contact the central gate that depletes the area between the two slits (serves also as plunger gate of the QD).

The collector current is related to the transmission probability from emitter to collector, $T_{EC}$, via the multiprobe conductance formula [11], $I_C = (2e^2/h)T_{EC}V_E$. As stressed above, the dominant contribution to $T_{EC}$ comes from the two direct paths, those going from *E* to *C* through the two slits (depicted by the two doted lines in Fig. 1 (a)), while longer paths reaching eventually the collector, resulting from multiple reflections from walls, are much less probable. Phase difference between the two direct paths is introduced via the AB effect. A magnetic flux, $\Phi$, threaded through the area enclosed by these two classical paths, *A*, results in an AB phase difference $\Delta\alpha = 2\pi\Phi/\Phi_0$ between the two interfering paths ($\Phi_0 = h/e$ is the flux quantum). Consequently, the collector current oscillates as a function of applied normal magnetic field with a period $\Delta B = \Phi_0 / A = 2.6$ mT, as seen in Fig. 2(a).

The suppression of AB oscillations due to the WP detector depends on the effect an electron dwelling in the QD has on the nearby detector and on the detector sensitivity. To study this we use the scheme seen in the inset of Fig. 2(b) [8], compose only of a QD and an adjacent QPC, fabricated on the same wafer. The control over the charge in the QD is done by a *plunger* gate, which affects only slightly the 'in' and 'out' barriers (formed by two QPCs) confining the QD. The resistance of each of these QPCs is adjusted to be greater than $h/2e^2$, forcing the QD deep in the Coulomb Blockade (CB) regime with well separated energy levels. Each CB peak in the conductance of the QD, scanned by the plunger gate voltage, $V_P$, (see Fig. 2(b)), is associated with adding a single electron to the QD. Tunneling between the QD and the QPC is negligibly small, however, due to their close proximity the transmission probability of the QPC detector, $T_d$ (related to its conductance, $g_d$, by the Landauer formula $g_d = (2e^2/h)T_d$) is affected by the potential of the nearby QD. As the plunger gate voltage is being scanned between two adjacent CB peaks, with fixed charge on the QD, the potential of the QD changes smoothly. When $V_p$ is being scanned across a CB peak and an electron is being added to the QD, a sharp rise of the potential, on the scale of the peak width, takes place. Consequently, the potential, and therefore $T_d$, are expected to exhibit a saw tooth - like oscillation, as indeed seen in the experimental results depicted in Fig. 2(b) (with $\Delta T_d$ reflecting the effect of an electron passing through the QD on the QPC). Fig 2 (c) shows $\Delta T_d$ as a function of $T_d$ (found by averaging over several CB peaks) with small $\Delta T_d$ near $T_d = 0$ and $T_d = 1$, an obvious consequence of approaching the conductance plateaus. This dependence will be used later. Note also that even though CB peaks are not observed for negative enough $V_P$ (due to the high resistance of the QD) the conductance of the QPC exhibits distinct saw tooth behavior, showing the existence of CB and conductance oscillations in the QD [8].

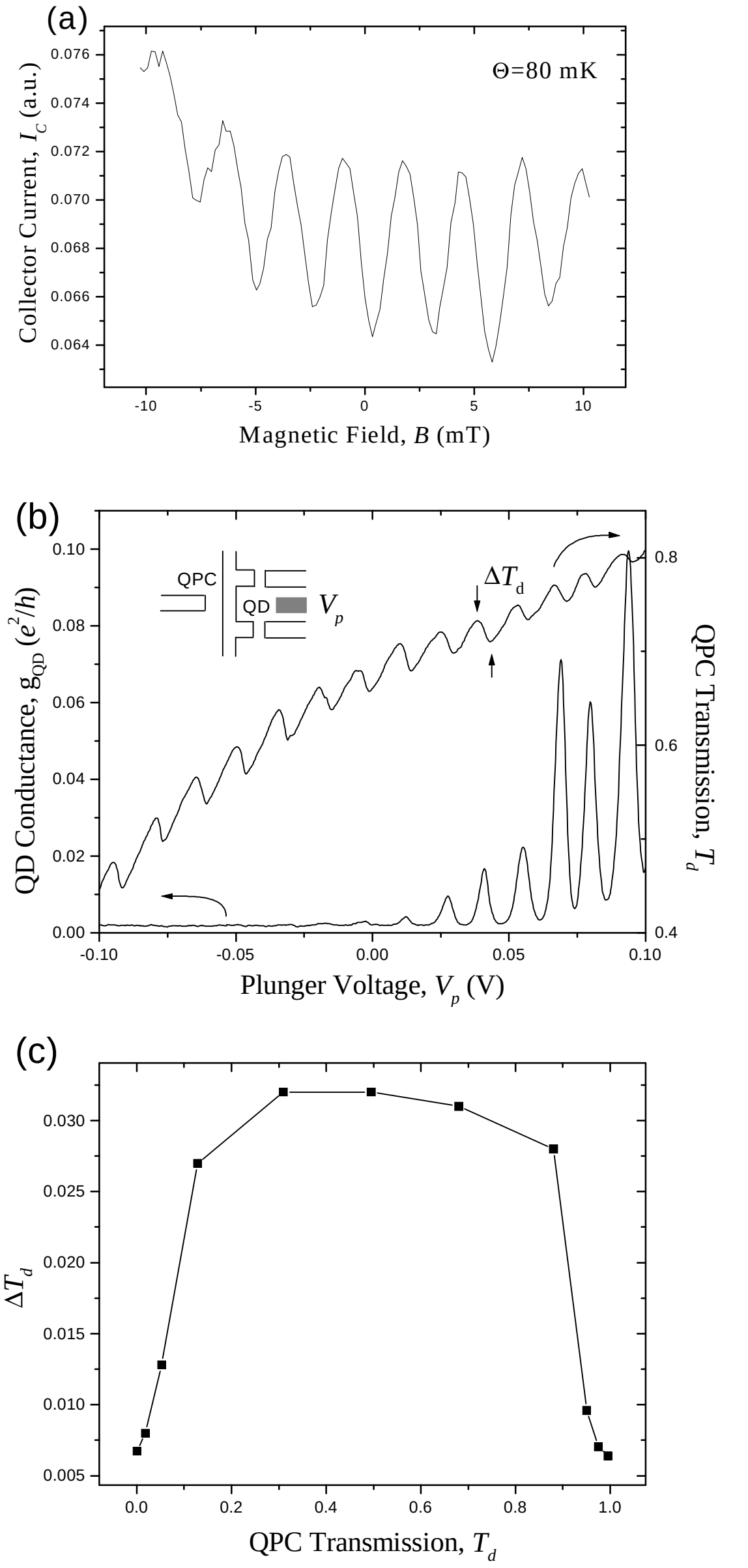

**Figure 2**: (a) AB oscillations in the collector current. (b) The conductance of the QD and that of the QPC detector as a function of the plunger gate voltage, $V_p$. The inset shows schematically the coupled structure. (c) The measured induced average change in the transmission probability of the QPC detector, $\Delta T_d$, due to charging the QD as a function of $T_d$.

What is the expected quantitative dephasing induced by the WP detector ? This problem had been treated recently by Aleiner, Wingreen and Meir [12], Levinson [13], and Gurvitz [14]. While Aleiner *et al.* looked on the effect of the QD on the detector, Levinson approached the problem from the opposite way; and indeed both reached similar results. On the other hand, Gurvitz's approach, based on quantum rate equations, led to a different result. A generalization of this problem and its connection to the theory of dephasing was discussed very recently by Imry [15]. We present here a simplified theoretical treatment that may provide a more intuitive picture of the dephasing process due to the WP detector.

Following Ref. [4] we write the entangled wave function of the whole system (interferometer + detector) as:

$$\left|\psi\right\rangle = \left|\varphi_l\right\rangle_e \otimes \left|\chi_l\right\rangle_d + e^{i\Delta\alpha}\left|\varphi_r\right\rangle_e \otimes \left|\chi_r\right\rangle_d \quad ,(1)$$

where $\left|\varphi_l\right\rangle_e$ ($\left|\varphi_r\right\rangle_e$) is the electronic partial wave associated with electron in the left (right) path, $\left|\chi_l\right\rangle_d$ ($\left|\chi_r\right\rangle_d$) represents the state of the detector coupled to the left (right) partial electronic wave, and $e^{i\Delta\alpha}$ is the phase shift between the two paths (in our case AB phase).

The probability to find the electron at the collector, $T_{EC}$, is found by summing (tracing out) over all possible states of the detector

(since the collector is sensitive only to the electron position, regardless the state of the detector). Assuming $|\chi_l\rangle_d$ and $|\chi_r\rangle_d$ are normalized one finds:

$$T_{EC} = \left|{}_e\langle\varphi_l|\mathbf{r}_C\rangle_e\right|^2 + \left|{}_e\langle\varphi_r|\mathbf{r}_C\rangle_e\right|^2 + 2\,\mathrm{Re}\left[e^{i\Delta\alpha}\,{}_e\langle\varphi_l|\mathbf{r}_C\rangle_e \cdot {}_e\langle\mathbf{r}_C|\varphi_r\rangle_e \cdot {}_d\langle\chi_r|\chi_l\rangle_d\right], \quad (2)$$

where $|\mathbf{r}_C\rangle_e$ represents the state of an electron at the detector. Therefore, the visibility of the interference pattern is given by: $v = v_0 v_d$, where:

$$v_0 = \frac{2\left|{}_e\langle\varphi_l|\mathbf{r}_C\rangle_e \cdot {}_e\langle\mathbf{r}_C|\varphi_r\rangle_e\right|}{\left|{}_e\langle\varphi_l|\mathbf{r}_C\rangle_e\right|^2 + \left|{}_e\langle\varphi_r|\mathbf{r}_C\rangle_e\right|^2}, \quad (3.1)$$

$$v_d = \left|{}_d\langle\chi_r|\chi_l\rangle_d\right| \quad . \quad (3.2)$$

The fact that $v_0$ can be smaller than 1 suggests that there is some *a priori* WP information even without the detector [16]. This is only possible for an asymmetric interferometer, e. g., one path is more probable than the other path. Similarly, $v_d$ represents the dephasing due to WP information obtained by the detector [16]. Note that after the interaction between the interfering electron and the environment is over, both states, $|\chi_l\rangle_d$ and $|\chi_r\rangle_d$, evolve according to the same unitary time evolution operator. Consequently, $v_d$ is time independent, regardless internal interactions of the environment. Hence, a question such as 'how the environment is measured ?' (if at all) is completely irrelevant for dephasing [15].

The effect of the electron dwelling in the QD - slit on the nearby QPC - detector leads to modified transmission and reflection amplitudes of the QPC, $t_r$ and $r_r$. Similarly, the amplitudes $t_l$ and $r_l$ are associated with an electron going through the left slit (being further away it affects the detector only very slightly). Using current conservation, time reversal symmetry (valid at $B=0$), and assuming that the barrier formed by the QPC detector is symmetric lead to $|t_i|^2 + |r_i|^2 = 1$ and $\mathrm{Re}(t_i r_i^*) = 0$. Thus, the transmission and reflection amplitudes may be written in the following way:

$$\begin{aligned} t_l &= \cos\theta_l \exp(i\varphi_l) \\ r_l &= i\sin\theta_l \exp(i\varphi_l) \\ t_r &= \cos\theta_r \exp(i\varphi_r) \\ r_r &= i\sin\theta_r \exp(i\varphi_r) \end{aligned} ,$$

where $\theta_l, \varphi_l, \theta_r, \varphi_r$ are real. The possible states of the detector can be described by using a basis of incoming and outgoing single particle (SP) states from both sides of the detector. Let $|I(\varepsilon)\rangle$ be an incoming SP state with energy $\varepsilon$, and let $|O(\varepsilon;t,r)\rangle = t|O_t(\varepsilon)\rangle + r|O_r(\varepsilon)\rangle$ be the evolving SP outgoing state, composed of transmitted and reflected partial waves. The inner product (overlap) between these SP outgoing states, $v_d^0 \equiv \left|\langle O(\varepsilon;t_r,r_r)|O(\varepsilon;t_l,r_l)\rangle\right|$, that represents the contribution of each electron probing the detector to $v_d$, is given by:

$$v_d^0 = \left|t_r^* t_l + r_r^* r_l\right| = \left|\cos\theta_r \cos\theta_l + \sin\theta_r \sin\theta_l\right| = \left|\cos(\theta_r - \theta_l)\right| .$$

Since in the experiment the effect of an electron dwelling in the QD on the nearby QPC detector is small we may assume $\Delta\theta \equiv \theta_r - \theta_l << 1$. Introducing the transmission probability $T_d \equiv |t|^2 = \cos^2\theta$ we have $\Delta T_d = (-2)\cos\theta\sin\theta\Delta\theta$, therefore:

$$v_d^0 = 1 - \frac{(\Delta T_d)^2}{8T_d(1-T_d)} \quad . \tag{4}$$

Note that only those single particle states that are different in $|\chi_l\rangle_d$ and $|\chi_r\rangle_d$ will contribute to $v_d$. The number of these states, $N$, is the number of particles that probe the detector during the time $\tau_d$ an electron dwells in the QD. Thus, the total overlap, $v_d$, is related to the SP overlap, $v_d^0$, by the simple relation: $v_d = (v_d^0)^N$. Note that we assume that the transmission and reflection coefficients are energy independent on the energy scale of temperature and chemical potential difference between both reservoirs connected to the detector, $eV_d$.

The rate particles probe the detector at zero temperature is $2eV_d / h$. The dwell time of an electron in the QD at resonance is $\tau_d = h / 2\pi\Gamma$, where $\Gamma$ is the width of the resonant state in the QD. Therefore, $N = (1/\pi) eV_d / \Gamma$. For our experimental conditions $N(\Delta\theta)^2 << 1$, therefore, $v_d = (1 - (\Delta\theta)^2 / 2)^N$ can be expressed as:

$$v_d = 1 - \frac{1}{\pi} \frac{eV_d}{\Gamma} \frac{(\Delta T_d)^2}{8T_d(1-T_d)} \quad . \tag{5}$$

Note that our result for the case of zero temperature agrees with the much more elaborate calculations of [12] and [13], while the factor of $(1-T_d)$ is missing in the result of Ref. [14].

It is interesting to represent $v_d$ in more physical terms of the detector performance [12]. Let $N_t$ be the number of transmitted particles out of the total $N$ particles probing the detector. This binomial random variable has an expectation value $\langle N_t \rangle = NT_d$ and a standard deviation (leading to shot noise) $\sigma(N_t) = \sqrt{NT_d(1-T_d)}$ [17]. Therefore, the standard deviation in the estimation of $T_d$ from counting the number of transmitted particles is $\sigma(T_d) = \sqrt{T_d(1-T_d)/N}$. Thus, our result can be written as:

$$v_d = 1 - \frac{1}{8}\left(\frac{\Delta T_d}{\sigma(T_d)}\right)^2 \quad .$$

For a noisy detector, $\sigma(T_d) >> \Delta T_d$, the detector provides no WP information and $v_d \approx 1$; while for a quiet detector, $\sigma(T_d) << \Delta T_d$, one can determine, even if 'in principle', the path the electron takes and consequently the interference pattern is expected to diminish.

In the actual experiment we measure the visibility of the AB conductance oscillations when the QD - slit is being tuned to a CB conduction peak (using the central island as a plunger gate) and the QPC - detector is conducting with transmission $0 \le T_d \le 1$. Figure 3 (a) shows the transmission probability of the QPC detector, $T_d$, and below we show in Fig. 3 (b) the visibility for two values of $V_d$, namely, probing rate of the detector, as a function of $V_g$, the voltage applied to the right gate of the QPC detector (see Fig. 1 (a)) that controls $T_d$. For $V_d = 100\ \mu\text{V}$ the visibility peaks when $\Delta T_d$ is small near conductance plateaus (namely, near $T_d = 0$ and $T_d = 1$), and also near $T_d = 0.5$ when the quantum shot noise is maximal. In these cases the WP detector is least sensitive to the electron dwelling in the adjacent QD and WP information is difficult to determine. These features altogether disappear when $V_d$ is being reduced to $10\ \mu\text{V}$. Reducing the probing rate of the electrons reduces the sensitivity of the detector. This observation confirms that the features found for $V_d = 100\ \mu\text{V}$ are indeed due to dephasing, rather than related to undesirable electrostatic effect of $V_g$ on the QD.

To compare our results with theory we generalize our calculation for the case of finite temperature [18]. We find that $\nu_0$ depends on the temperature of the interferometer with a cross over between 'low' and 'high' temperature at $k_B\Theta \approx \Gamma$. On the other hand, $\nu_d$ depends mainly on the temperature of the detector with a cross over between 'low' and 'high' temperature at $k_B\Theta \approx eV_d$. Note that this temperature dependence is in general different from the one found in Ref. [12]. For the comparison with experiment, however, we use the results of [18] since they better agree with the measured visibility. The visibility for the case $V_d = 100\ \mu\text{V}$ is expected to be proportional to the zero temperature value of $\nu_d$ (Eq. (5)) since $eV_d >> k_B\Theta \approx 7\mu\text{eV}$. We use the measured $\Delta T_d$ in the calibration device (Fig. 2 (c)) and the value of $\Gamma = 0.5\ \mu\text{eV}$ as a fitting parameter. The calculated visibility, drawn as a solid line in Fig. 3 (b), exhibits a reasonable agreement with experiment, both qualitatively and quantitatively. The dependence of the visibility on drain source voltage, $V_d$ (measured in a different working regime than in Fig. 3 (b)) is given in Fig. 3 (c). For $eV_d >> k_B\Theta$ the dependence is linear as expected from Eq. (5), with a slope corresponding to $\Gamma = 0.7\ \mu\text{eV}$ (which is reasonable for the retuned QD and interferometer). The deviation from linear dependence near $V_d = 0$ can be accounted due to the finite temperature ($eV_d \approx k_B\Theta$) [18].

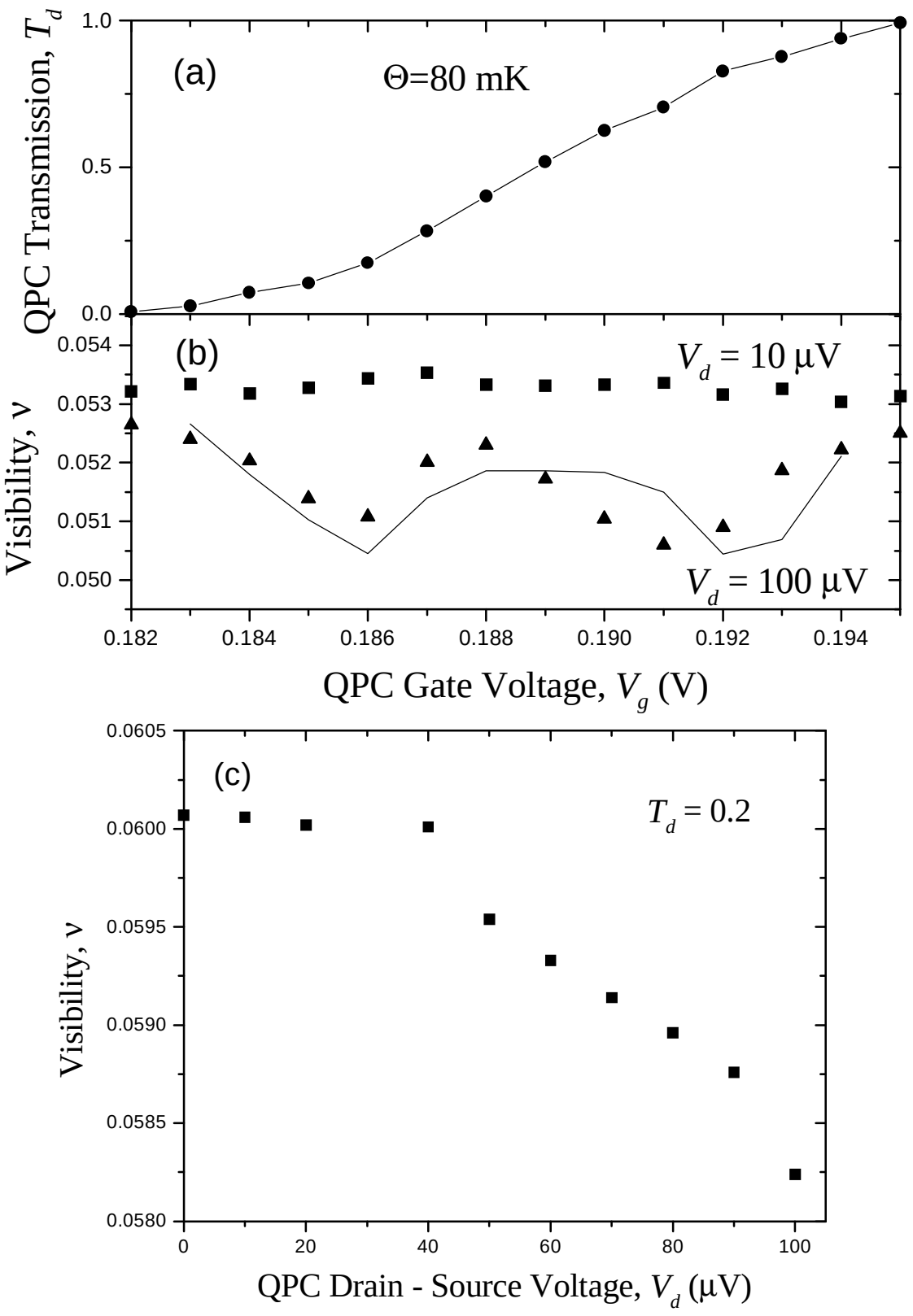

**Figure 3**: (a) The transmission probability of the QPC detector, $T_d$, as a function of the voltage applied to the right gate of the QPC detector, $V_g$. (b) The visibility of the AB oscillations as a function of $V_g$ for two values of the drain source voltage across the detector, $V_d$. (c) The visibility of the AB oscillations as a function of $V_d$ for a fixed $T_d = 0.2$.

Aside from demonstrating the principle of complementarity we believe that similar experimental setups with higher detector sensitivity may be used to study other fundamental problems in quantum mechanics. For example, looking at the current of the WP detector in the time domain may shed some light on the old and controversial issue of wave function reduction. Increasing the mutual coupling between detector and QD might open a way to fabricate a quantum bit (qubit) with possible applications in quantum computing.

We thank S. Gurvitz for presenting to us Ref. [9] that initiated the present work. We thank also for most valuable discussions with I. Imry, Y. Levinson, Y. Meir, A. Stern, and N. Wingreen. This work has been partly supported by a MINERVA grant and a MINERVA fellowship for one of us (R. S.).